\documentclass[lettersize,journal]{IEEEtran}
\usepackage{amsmath,amsfonts}
\usepackage{algorithmic}
\usepackage[ruled,linesnumbered]{algorithm2e}
\usepackage{array}
\usepackage[caption=false,font=normalsize,labelfont=sf,textfont=sf]{subfig}
\usepackage{textcomp}
\usepackage{stfloats}
\usepackage{url}
\usepackage{verbatim}
\usepackage{graphicx}
\usepackage{cite}
\usepackage{multicol}
\usepackage{multirow}
\usepackage{pfnote}
\usepackage{booktabs}
\usepackage{float}
\usepackage{color}
\begin{document}

\title{Generalizable Learning Reconstruction for Accelerating MR Imaging via Federated Neural Architecture Search}

\author{Ruoyou Wu, Cheng Li, Juan Zou, Shanshan Wang,~\IEEEmembership{Senior Member,~IEEE}
\thanks{Manuscript received  August 23, 2023. (\textit{Corresponding author: Shanshan Wang.})}
\thanks{Ruoyou Wu is with the Paul C. Lauterbur Research Center for Biomedical Imaging, Shenzhen Institute of Advanced Technology, Chinese Academy of Sciences, Shenzhen 518055, China, and with the Peng Cheng Laboratory, Shenzhen 518055, China, and also with the University of Chinese Academy of Sciences, Beijing 100049, China (e-mail: ry.wu@siat.ac.cn).}
\thanks{Cheng Li is with the Pual C. Lauterbur Research Center for Biomedical Imaging, Shenzhen Institute of Advanced Technology, Chinese Academy of Sciences, Shenzhen 518055, China (e-mail: cheng.li6@siat.ac.cn).}
\thanks{Juan Zou is with the School of Physics and Optoelectronics, Xiangtan University, Xiangtan 411105, China (e-mail: zjuan@xtu.edu.cn).}
\thanks{Shanshan Wang is with the Paul C. Lauterbur Research Center for Biomedical Imaging, Shenzhen Institute of Advanced Technology, Chinese Academy of Sciences, Shenzhen 518055, China, and also with the Peng Cheng Laboratory, Shenzhen 518055, China (e-mail: sophiasswang@hotmail.com).}
}


\IEEEpubid{0000--0000/00\$00.00~\copyright~2023 IEEE}

\maketitle

\begin{abstract}
Heterogeneous data captured by different scanning devices and imaging protocols can affect the generalization performance of the deep learning magnetic resonance (MR) reconstruction model. While a centralized training model is effective in mitigating this problem, it raises concerns about privacy protection. Federated learning is a distributed training paradigm that can utilize multi-institutional data for collaborative training without sharing data. However, existing federated learning MR image reconstruction methods rely on models designed manually by experts, which are complex and computational expensive, suffering from performance degradation when facing heterogeneous data distributions. In addition, these methods give inadequate consideration to fairness issues, namely, ensuring that the model's training does not introduce bias towards any specific dataset's distribution. To this end, this paper proposes a generalizable federated neural architecture search framework for accelerating MR imaging (GAutoMRI). Specifically, automatic neural architecture search is investigated for effective and efficient neural network representation learning of MR images from different centers. Furthermore, we design a fairness adjustment approach that can enable the model to learn features fairly from inconsistent distributions of different devices and centers, and thus enforce the model generalize to the unseen center. Extensive experiments show that our proposed GAutoMRI has better performances and generalization ability compared with six state-of-the-art federated learning methods. Moreover, the GAutoMRI model is significantly more lightweight, making it an efficient choice for MR image reconstruction tasks. The code will be made available at https://github.com/ternencewu123/GAutoMRI.

\end{abstract}


\section{Introduction}
\IEEEPARstart{M}{agnetic} resonance imaging (MRI) is a non-invasive imaging modality that can effectively assess the structure, morphology, and function of soft tissues. It plays a crucial role in clinical diagnosis and disease evaluation due to its ability to provide detailed and comprehensive information. However, acquiring fully-sampled $k$-space data is usually time-consuming due to the physical limitations of the scanning device\cite{korkmaz2022unsupervised}. One possible way to improve the imaging speed is to reconstruct high-quality images from undersampled data\cite{sandino2020compressed}. Recently, deep learning has shown great potential in accelerating MR image reconstruction\cite{wang2016accelerating,hammernik2018learning,mardani2018deep,yang2018admm,quan2018compressed,aggarwal2018modl,zhu2018image,qin2018convolutional,han2018deep,akccakaya2019scan,wang2021deep,lee2018deep, wang2020deepcomplexmri,eo2018kiki}. Deep learning offers the advantage of fast online reconstruction through offline learning from large datasets. However, due to factors like varying scanning devices, imaging protocols, and other discrepancies, the distribution of data among different centers can become highly heterogeneous. As a consequence, generalizing a model trained in one center to other centers becomes a challenging task, and ensuring the stability of the model becomes equally difficult. This could lead to biased performance at different centers. One conventional approach is to aggregate data from multiple centers for centralized learning\cite{liang2020deep}. However, collecting data from multiple centers can lead to significant resource consumption and may leak patient privacy\cite{kaissis2020secure, kaissis2021end}.

\IEEEpubidadjcol  

To address this limitation, federated learning emerges as a promising framework that allows multi-institutional collaborative learning while protecting patient privacy\cite{li2020federated, mcmahan2017fedavg,li2020fedprox,li2021moon,liu2021feddg,zhang2023fedala,guo2021fl-mrcm,feng2022fedmri,elmas2022federated,wang2021federated}. In federated learning, each client independently trains a local model using its computational resources and private data. Subsequently, the server collects these locally trained models from all clients and employs an aggregation approach to get the server model. Finally, the server distributes the aggregated model to the clients for the subsequent communication rounds\cite{mcmahan2017fedavg}.
The classical federated learning method FedAvg\cite{mcmahan2017fedavg} is effective for client model aggregation with the same distribution, but in MRI, the data distribution of different clients may present heterogeneity caused by different tissue contrasts, imaging protocols, and other factors. This may cause models trained by federated learning to suffer performance degradation. To this end, several federated learning MRI methods have been proposed to alleviate this problem. In\cite{guo2021fl-mrcm}, the similarity of latent-space representations is improved by repeatedly aligning the source and target sites. In \cite{feng2022fedmri}, client specificity is maintained by dividing the reconstruction model into a globally shared encoder and a locally personalized decoder. While existing federated learning MRI reconstruction methods have achieved promising performance, the reconstruction models of these methods typically utilize hand-designed network architectures, which are often complex and require more computational resources. Additionally, these methods also rarely consider the fairness of the algorithm, which involve an obvious performance gap of the model across different centers, especially in the case of heterogeneous data distributions. This makes it difficult to generalize the trained model to other centers\cite{wang2021federated}.

Neural Architecture Search (NAS)\cite{elsken2019neural} can achieve better performance by automatically searching for the optimal network architecture while reducing the consumption of computational resources. A typical NAS architecture consists of three parts: search space, search approach, and evaluation approach. Recently, there have been some works on utilizing NAS alone for MR image reconstruction\cite{yan2020neural,huang2020enhanced}. All of these methods have yielded improved results, achieving better performance with fewer model parameters when compared to the existing hand-designed models. However, these methods don't consider heterogeneous data distributions from different centers.

To address the above mentioned challenges,  we propose a generalizable federated neural architecture search framework for accelerating MR imaging. Our main contributions can be summarized as follows:

1) A generalizable federated neural architecture search framework (GAutoMRI) is proposed for accurate under-sampled MR image reconstruction with lightweight models, using multi-institutional data in a privacy-preserving manner.

2) A fairness adjustment approach is designed to enhance the learning capabilities of the model across different centers, enabling the model to better generalize to both in-distribution and out-of-distribution data.

3) Extensive experimental validations have been performed on three public and one in-house data set with different undersampling patterns, imaging modalities, and scanning devices. Results show that our proposed GAutoMRI has better performances
and generalization capability compared with six state-of-the-art federated learning methods with significantly lightweight models.

\section{Related Work}
\subsection{MR Image Reconstruction}
Reconstructing high-quality MR images from undersampling $k$-space data is a typical ill-posed inverse problem. Deep learning-based methods learn the mapping relationship from undersampled to fully-sampled data offline and then enable fast online reconstruction\cite{wang2016accelerating}. These methods can avoid manual fine-tuning of optimization parameters. Deep learning-based MR image reconstruction methods can be broadly categorized into data-driven\cite{mardani2018deep,quan2018compressed,zhu2018image,han2018deep,akccakaya2019scan,wang2021deep,lee2018deep,wang2020deepcomplexmri,eo2018kiki} and model-driven approaches\cite{hammernik2018learning,yang2018admm,aggarwal2018modl,qin2018convolutional}. Data-driven methods typically rely on large amounts of data and directly learn the mapping relationship from undersampled data to fully-sampled data. However, the networks could be totally black-box. Model-driven approaches formulate an optimization problem and then unfold the iterative optimization algorithm into the network. The network architecture is constructed based on iterative steps, leading to a slightly longer reconstruction time. While these methods have made certain contributions to the development of MR image reconstruction, they all require the collection of large amounts of data for centralized training. This undoubtedly increases labor costs and resource consumption, and it also brings a serious issue---data privacy leakage. We utilize a distributed collaborative training scheme to effectively avoid the problems of data collection and patient privacy leakage.

\subsection{Federated Learning}
Federated learning is a distributed collaborative training paradigm that protects data privacy by training shared model without sharing data\cite{li2020federated,mcmahan2017fedavg,li2020fedprox,li2021moon,liu2021feddg,zhang2023fedala,guo2021fl-mrcm,feng2022fedmri,elmas2022federated,wang2021federated}. One of the most classical federated learning methods is FedAvg\cite{mcmahan2017fedavg}, which obtains the global model through an average aggregation method, and it obtains a relatively better performance when the data distributions of different centers are similar. However, in real-world scenarios, there is often statistical heterogeneity among different centers, and the traditional federated learning methods are difficult to adapt to such requirements. Some works start to reconcile the differences between global and local models \cite{li2020fedprox,li2021moon}, aiming to obtain a global model with better performance. For example, Li et al.,\cite{li2020fedprox} facilitate the learning of the global model by imposing an additional constraint on the client's objective function. In addition, there are also some proposed personalized federated learning methods to preserve the specificity of local data. For example, Zhang et al.,\cite{zhang2023fedala} proposed a local adaptive aggregation approach to improve the performance of local clients by adaptively aggregating global and local models. As for magnetic resonance imaging, the problem of data heterogeneity may also exist in different centers due to scanning devices, imaging protocols, and other issues\cite{knoll2020fastmri}. Some methods start to solve the problems of feature drift and domain generalization. For example, Guo et al.,\cite{guo2021fl-mrcm} reduces the data heterogeneity of clients by repeatedly aligning the hidden feature distributions between the source and target domains. It selects a target site in each communication round and then aligns it with the representation of the remaining sites, which adds more communication consumption. In addition, Feng et al.,\cite{feng2022fedmri} effectively preserve local client-specific features by dividing the reconstruction model into a globally shared encoder and a locally personalized decoder. These approaches effectively mitigate the privacy leakage problem, but their model architectures are hand-designed by experts, which are often complex and require more computational resources. We utilize an automatically differentiable architecture search method to obtain an optimal model architecture, effectively reducing the number of model parameters and the need for computational resources.

\subsection{Neural Architecture Search}
Neural architecture search can obtain optimal network architectures by employing an automated network architecture search approach while keeping fewer model parameters\cite{elsken2019neural}. Based on different search strategies, neural architecture search algorithms can be broadly categorized into three types: reinforcement learning (RL)-based\cite{bello2017neural,pham2018efficient}, evolutionary algorithm (EA)-based\cite{yang2020cars,sun2020automatically}, and gradient-based (GD) methods\cite{liu2018darts,he2020milenas}. Reinforcement learning-based and evolutionary algorithm-based methods are gradient-free methods, in contrast to gradient-based methods that tend to be faster in the searching process and consume slightly fewer computational resources\cite{he2020milenas}. DARTS is the most classical gradient-based method, which scales the selection of candidate operations into continuous variables and then utilizes efficient gradient back-propagation for learning, effectively reducing the search time\cite{liu2018darts}. In addition, there is also some NAS-based work in MR image reconstruction\cite{yan2020neural,huang2020enhanced}. For example, Yan et al., \cite{yan2020neural} used NAS for compressed sensing MR image reconstruction, achieving enhanced performance. These NAS-based MR image reconstruction methods are used for centralized training scenarios but have not yet been explored in federated learning scenarios. In this paper, we utilize federated learning and neural architecture search for MR image reconstruction and improve the generalization performance for both in-distribution and out-of-distribution data through a fairness adjustment approach.

\section{Methodology}
\subsection{DL-based MR Image Reconstruction}
The goal of MR image reconstruction is to reconstruct a high-quality image $\mathbf {x}\in \mathbb{C}^{N}(M<N)$ from undersampled $k$-space data $\mathbf {b}\in \mathbb{C}^{M}$, formulated as:

\begin{equation}
\mathbf{b} = \mathbf{Ax}+\mathbf{\epsilon }  
\label{equ1}
\end{equation}
where $\mathbf{A}\in\mathbb{C}^{M\times N} $ denotes the encoding matrix and $\mathbf{\epsilon }\in\mathbb{C}^{M} $ denotes the measurement noise. Solving $\mathbf{x}$ from Eq. \eqref{equ1} is an ill-posed inverse problem that can be expressed as the following optimization problem:

\begin{equation}
    \min_{\mathbf{x}}\frac{1}{2}\left \| \mathbf{Ax}-\mathbf{b}   \right \|+\lambda R(\mathbf{x} )
\label{equ2}
\end{equation}
where the first term is the data fidelity, the second term is the regularization term, and $\lambda$ is the regularization parameter used to balance the tradeoff between the data fidelity term and the regularization term. Based on the method of model unfolding\cite{aggarwal2018modl}, Eq. \eqref{equ2} can be transformed into the following alternating iteration process:

\begin{equation}
\left\{
    \begin{array}{lr}
    \mathbf{z}^{j}=f_{\theta}\left (\mathbf{x}^{j}  \right ) &  \\
    \mathbf{x}^{j+1}=\left (\mathbf{A}^{H}\mathbf{A}+\lambda \mathbf{I}  \right ) ^{-1}\left (\mathbf{A}^{H}\mathbf{b} + \lambda \mathbf{z}^{j}\right )
    \end{array}
\right.
\label{equ3}
\end{equation}
where $f_{\theta}$ denotes the deep neural network used for denoising, $\mathbf{A}^{H}$ denotes the conjugate operator of $\mathbf{A}$, and $j$ denotes the iterative control variable, $\lambda$ is learnable.

\subsection{Details of GAutoMRI}
\begin{figure*}[t]
    \centering
    \setlength{\abovecaptionskip}{0.cm}
    \includegraphics[width=14cm, height=8cm, keepaspectratio]{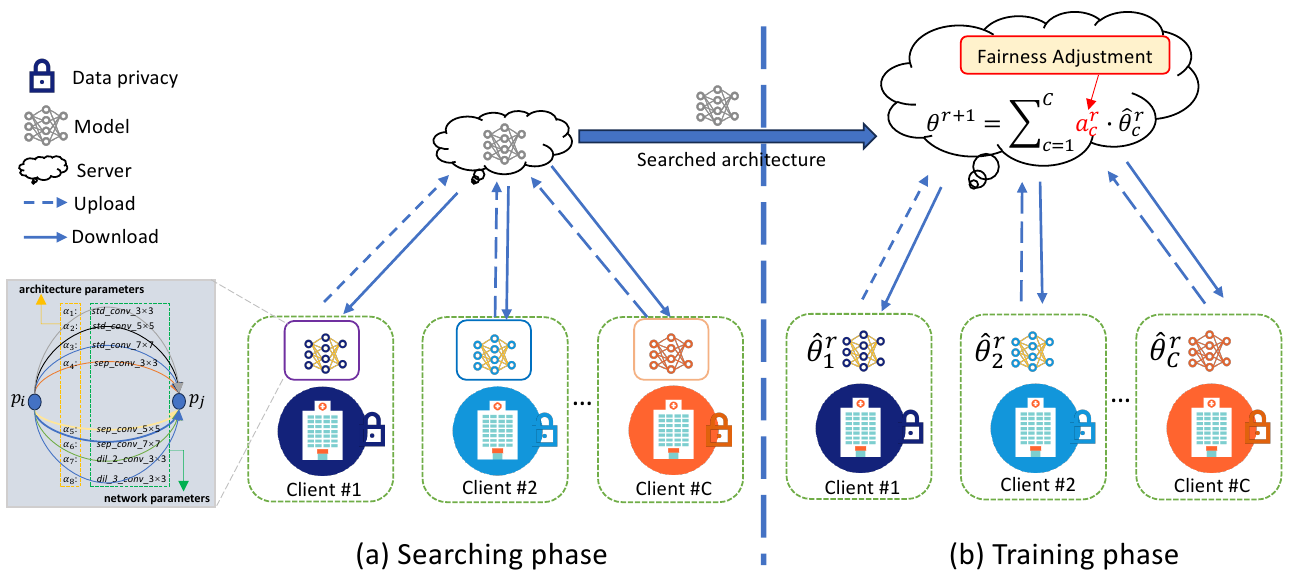}
    \caption{The overall architecture of our proposed GAutoMRI. It consists of two phases: (a) Searching phase. The aim of this phase is to obtain the optimal architecture using the federated neural architecture search method and use the searched architecture for the training phase. In this paper, we utilize the approach outlined in \cite{he2020towards} to achieve this, while adjusting the search space to suit our specific MR image reconstruction task. (b) Training phase. The optimal architecture obtained in the searching phase is utilized as the base model. In addition, we introduce a fairness adjustment approach to improve the generalization of the global model over out-of-distribution data, as detailed in III.C.
    }
    \label{fig1}
\end{figure*}

The overall architecture of GAutoMRI is shown in Fig. \ref{fig1}, which consists of two phases: the searching phase and the training phase. In the searching phase, we adjusted the search space according to the reconstruction task, details of which can be found in III.B.2 and Fig. \ref{fig2}, and the whole search process draws on the method in \cite{he2020towards} to obtain the optimal network architecture. In the training phase, the architecture obtained in the searching phase is employed as the base model. In addition, we introduce a fairness adjustment approach to improve the generalization of the global model over out-of-distribution data. The specific search process and fairness adjustment approach will be described later.

\subsubsection{Preliminaries}
In the classical federated learning setting, the model architecture is fixed. It is assumed that there are $C$ clients or hospitals, each with a dataset $D_{c}=\{(x_{c}^{i}, b_{c}^{i})\}_{i=1}^{N_{c}}$, which is non IID. The goal of federated learning is to learn a global model on dataset $D = \{D_{1}, D_{2},..., D_{C}\}$ in a distributed training manner, which can be expressed as:

\begin{equation}
\min_{\theta}\mathfrak{L(\theta)}=\sum_{c=1}^{C}\frac{N_{c}}{N}\{\mathbb{E}_{(\mathbf{x}_{c},\mathbf{b}_{c})\in \mathcal{D}_{c}}\ell_{c}(f(\mathbf{b}_{c};\theta))\}
\label{equ4}
\end{equation}
where $N_{c}$ and $N$ denote the number of samples for client $c$ and the total number of samples, respectively; $\theta$ denotes network parameters; $ \ell_{c}(\cdot)$ denotes the local loss on client $c$. For federated neural architecture search, it is necessary to optimize both model parameters $\theta$ and architecture parameters $\alpha$. Formally, the objective function is defined as:

\begin{equation}
    \min_{\theta, \alpha }\sum_{c=1}^{C}\frac{N_{c}}{N}\{ \mathbb{E}_{(\mathbf{x}_{c},\mathbf{b}_{c})\in \mathcal{D}_{c}}\ell_{c}(f(\mathbf{b}_{c};\theta,\alpha))\} 
\label{equ5}
\end{equation}
Our goal is to search the optimal architecture parameter $\alpha$ and the corresponding model parameters $\theta$ by utilizing a differentiable architecture search method, making it fit the heterogeneous data distribution, thereby improving the generalization performance of out-of-distribution data.

\subsubsection{Search Space}
\begin{figure}[t]
    \centering
    \setlength{\abovecaptionskip}{0.cm}
    \includegraphics[width=7cm, height=7cm, keepaspectratio]{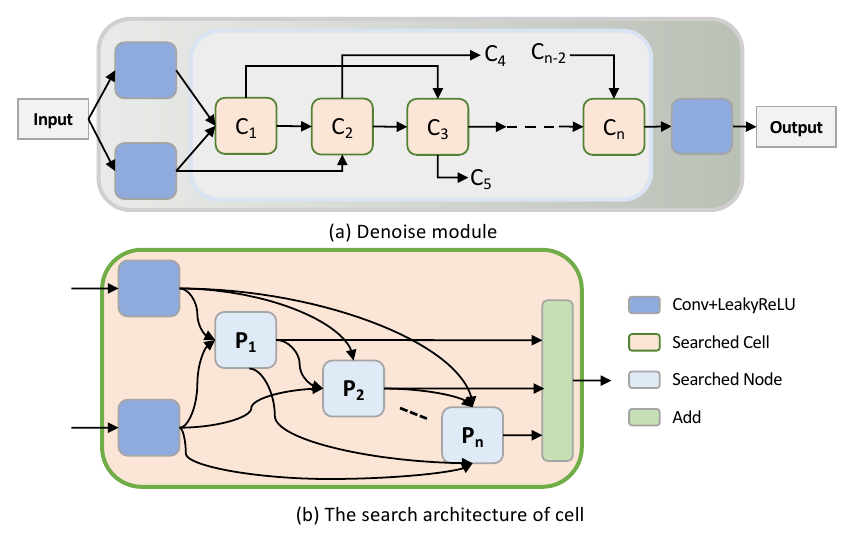}
    \caption{The overall structure of search space. (a) The structure of denoiser module $f_{\theta}(\cdot)$ in our reconstruction network. The searched cells are stacked to form our basic architecture, with each cell connected to the two consecutive cells that follow it. (b) The search architecture of the cell. $P_{n}$ denotes the node in the cell. 
    }
    \label{fig2}
\end{figure}

Typically, a NAS method consists of three consecutive parts: search space, search approach, and performance evaluation approach\cite{elsken2019neural}. Our search space follows the mixed-operation search space defined in DARTS\cite{liu2018darts}, as shown in Fig. \ref{fig2}. We designed our candidate set of operations, denoted as $\mathcal{O}$, based on the number of parameters and test performance of the target architecture, aiming to make it more suitable to our reconstruction task. The set includes: 1) Standard convolution, which extracts multi-scale information using convolution kernels of different sizes: $std\_conv\_3\times3$, $std\_conv\_5\times5$, $std\_conv\_7\times7$; 2) Depthwise separable convolution, reducing parameters and computation while improving feature extraction and representation ability: $sep\_conv\_3\times3$, $sep\_conv\_5\times5$ and $sep\_conv\_7\times7$; 3) Dilated convolution, enlarging receptive field and reducing the number of parameters: $dil\_2\_conv\_3\times3$ and  $dil\_3\_conv\_3\times3$. An edge between two nodes in Fig. \ref{fig2}(b) represents one of these candidate operations. Inside the cell, to make the search space continuous by relaxing the categorical candidate operations between two nodes, mixed operations using the $softmax$ function over all candidate operations are performed:

\begin{equation}
      \bar{\phi}^{(i,j)}(x)=\sum_{o=1}^{|\mathcal{O} | }\frac{exp(\alpha_{o}^{(i,j)} )}{ {\textstyle \sum_{o^{'}=1}^{|\mathcal{O}|}}exp(\alpha_{o^{'}}^{(i,j)})}\phi_{o}^{(i,j)}(x)
   \label{equ6}
\end{equation}
where $\bar{\phi}^{(i,j)}(x)$ denotes the mixed operation for a pair of nodes $(i,j)$, $|\mathcal{O}|$ denotes the number of candidate operations, $\alpha_{o}^{(i,j)}$ denotes the weight of the $o^{th}$ operation for a pair of node $(i,j)$, and $\phi_{o}^{(i,j)}(x)$ denotes the $o^{th}$ operation for a pair of node $(i,j)$. The aim of architecture search is to learn the optimal encoding $\alpha=\{\alpha^{(i,j)}\}$ of the architecture.

\begin{algorithm}[ht]
 \caption{\textbf{GAutoMRI-Searcher} follows in \cite{he2020towards}}
 \label{alg1}
 \SetKwInOut{Input}{Input}\SetKwInOut{Output}{Output}
 \SetKwProg{Server}{ServerExecution}{:}{}
 \SetKwProg{Local}{LocalSearch}{}{}
 
  \Input{The datasets from $C$ clients: $\mathcal{D}_{1}, \mathcal{D}_{2},\cdots,\mathcal{D}_{C}= \{ (\mathbf{b}^{i}, \mathbf{x}^{i})\mid i=1,\cdots ,N_{C} \}  $; communication rounds: $R$; local epochs: $Z$; learning rate for $\alpha$: $\eta_{\alpha}$; learning rate for $\theta$: $\eta_{\theta}$ total sample size: $N$;
  tradeoff coefficient: $\lambda$ \;}
  
  \BlankLine
  \Server{}
  {
  Initialize global parameters $\alpha^{0}$ and $\theta^{0}$\;
  \For{$r=1,2,\cdots, R$}{
    \For{$c=1,2,\cdots, C$}{
     Get global parameters $\alpha_{c}^{r}$, $\theta_{c}^{r}$ $\gets$ $\alpha^{r}$, $\theta^{r}$ \;
     $\hat{\alpha}_{c}^{r}$, $\hat{\theta}_{c}^{r}$ $\gets$ \textbf{LocalSearch}(c, $\alpha_{c}^{r}$, $\theta_{c}^{r}$)\;
    }
    $\alpha^{r+1} \gets {\textstyle \sum_{c=1}^{C}\frac{N_{c}}{N}\hat{\alpha}_{c}^{r}} $ \;
    $\theta^{r+1} \gets {\textstyle \sum_{c=1}^{C}\frac{N_{c}}{N}\hat{\theta}_{c}^{r}} $ \;
  }
  return $\alpha^{r+1}$ and $\theta^{r+1}$ \;
  }
  \Local{($c,\theta, \alpha$):}{
  \For{$z=1,2,\cdots, Z$}{
  Get the updated parameters $\bigtriangledown_{\alpha}\mathfrak{L}_{tr} (\theta_{c}^{r,z}, \alpha_{c}^{r,z} )$ and $\bigtriangledown_{\alpha}\mathfrak{L}_{val} (\theta_{c}^{r,z}, \alpha_{c}^{r,z} ) $ \;
  Update $\hat{\alpha}_{c}^{r,z+1}\gets \alpha_{c}^{r,z}-\eta_{\alpha}(\bigtriangledown_{\alpha}\mathfrak{L}_{tr} (\theta_{c}^{r,z}, \alpha_{c}^{r,z} )+$ \\ 
$\lambda\bigtriangledown_{\alpha}\mathfrak{L}_{val} (\theta_{c}^{r,z}, \alpha_{c}^{r,z} ))$ \;
  Get the updated parameters $\bigtriangledown_{\theta}\mathfrak{L}_{tr} (\theta_{c}^{r,z}, \alpha_{c}^{r,z} )$ \;
  Update $\hat{\theta}_{c}^{r,z+1}\gets \theta_{c}^{r,z}-\eta_{\theta}(\bigtriangledown_{\theta}\mathfrak{L}_{tr} (\theta_{c}^{r,z}, \alpha_{c}^{r,z} ))$ \;
  }
  return $\hat{\alpha}_{c}^{r}$ and $\hat{ \theta}_{c}^{r}$ \;
  }
\end{algorithm}

\subsubsection{Search Process}
During the search process, to optimize the objective function in Eq. \eqref{equ5}, we refer to the method in \cite{he2020towards} to find the optimal architecture with a distributed neural architecture search algorithm. The details of the search process are shown in Algorithm. \ref{alg1}. First, the initialized global architecture parameters $\alpha^{0}$ and model parameters $\theta^{0}$ are distributed to each local client; then each client updates the architecture parameters $\alpha_{c}^{r}$ and model parameters $\theta_{c}^{r}$ using the local private data as follows:

\begin{equation}
    \left\{\begin{array}{lr}
\hat{\alpha}_{c}^{r,z+1}\gets \alpha_{c}^{r,z}-\eta_{\alpha}(\bigtriangledown_{\alpha}\mathfrak{L}_{tr} (\theta_{c}^{r,z}, \alpha_{c}^{r,z} )+\\
\lambda \bigtriangledown_{\alpha}\mathfrak{L}_{val} (\theta_{c}^{r,z}, \alpha_{c}^{r,z} )) 
 \\
\hat{\theta}_{c}^{r,z+1}\gets \theta_{c}^{r,z}-\eta_{\theta}(\bigtriangledown_{\theta}\mathfrak{L}_{tr} (\theta_{c}^{r,z}, \alpha_{c}^{r,z} ))
\end{array}\right.
\label{equ7}
\end{equation}
where $r$ and $z$ denote global communication rounds and local training epochs, respectively. $\mathfrak{L}_{tr} (\theta_{c}^{r,z}, \alpha_{c}^{r,z} )$ and $\mathfrak{L}_{val} (\theta_{c}^{r,z}, \alpha_{c}^{r,z} )$ denote the local training loss and validation loss with respect to the architecture parameters $\alpha_{c}^{r,z}$ and model parameters $\theta_{c}^{r,z}$, respectively. $\eta_{\alpha}$ and $\eta_{\theta}$ denote the learning rate of architecture parameters $\alpha$ and model parameters $\theta$, respectively. 

After completing the local training, each client sends the architecture parameters $\hat{\alpha}_{c}^{r}$ and model parameters $\hat{\theta}_{c}^{r}$ to the server. The server receives the parameters from each client and performs the aggregation operation as follows:

\begin{equation}
     \left\{\begin{array}{lr}
\alpha^{r+1}\gets \sum_{c=1}^{C}\frac{N_{c}}{N}\hat{\alpha}_{c}^{r}
 \\
\theta^{r+1}\gets \sum_{c=1}^{C}\frac{N_{c}}{N}\hat{\theta}_{c}^{r}
\end{array}\right.
\label{equ8}
\end{equation}
where $\alpha^{r+1}$ and $\theta^{r+1}$ denote the global architecture parameters and model parameters for the $r^{th}$ communication round, respectively. Then, the server distributes the global parameters $\alpha^{r+1}$ and $\theta^{r+1}$ to each client for the next communication round. After $R$ communication rounds, we can obtain the optimal architecture parameters and model parameters without sharing local private data.

\begin{algorithm}[ht]
 \caption{\textbf{GAutoMRI-Trainer}}
 \label{alg2}
  \SetKwInOut{Input}{Input}\SetKwInOut{Output}{Output}
  \SetKwProg{Server}{ServerExecution}{:}{}
  \SetKwProg{Local}{LocalSearch}{}{}

  \Input{The datasets from $C$ clients: $\mathcal{D}_{1}, \mathcal{D}_{2},\cdots,\mathcal{D}_{C}= \{ (\mathbf{b}^{i}, \mathbf{x}^{i})\mid i=1,\cdots ,N_{C} \}  $; communication rounds: $R$; local epochs: $Z$; learning rate for $\theta$: $\eta_{\theta}$; initialized aggregation weights: $a_{c}^{0}=\frac{1}{C} $ ; Incremental control parameters: $\gamma$\;}
 
  \BlankLine
  \Server{}{
  Initialize global parameters $\theta^{0}$ \;
  \For{$r=1,2,\cdots, R$}{
    \For{$c=1,2,\cdots, C$}{
      Get global parameters $\theta_{c}^{r} \gets \theta^{r}$\;
      $\mathcal{G}_{D_{c}}(\theta^{r})$, $\hat{\theta}_{c}^{r}$ $\gets$ \textbf{LocalSearch}($c, \theta_{c}^{r}$)\;
    }
    Update the aggregation weights $a_{c}^{r}$ based on Eq. \eqref{equ10} and Eq. \eqref{equ11}\;
    
    Aggregate $\theta^{r+1} \gets \sum_{c=1}^{C}a_{c}^{r}\cdot \hat{\theta}_{c}^{r}$\;
  }
  return $\theta^{r+1}$
  }
  \Local{(c, $\theta$):}{
  Get the updated parameters $\theta^{r}$\;
  Compute $\mathcal{G}_{D_{c}}(\theta^{r})$ based on Eq. \eqref{equ9}\;
  \For{$z=1,2,\cdots, Z$}{
  Get the updated gradients $\bigtriangledown_{\theta}\mathfrak{L}_{tr}(\mathbf{x}_{c},\mathbf{b}_{c};\theta_{c}^{r,z})$\;
  
  Update $\hat{\theta}_{c}^{r,z+1}\gets \theta_{c}^{r,z}-\eta_{\theta}(\bigtriangledown_{\theta}\mathfrak{L}_{tr}(\mathbf{x}_{c},\mathbf{b}_{c};\theta_{c}^{r,z}))$\;
  }
  return $\mathcal{G}_{D_{c}}(\theta^{r})$ and $\hat{\theta}_{c}^{r}$ \
  }
\end{algorithm}

\subsection{Fairness Adjustment for FL}
After the searching phase, we can obtain an optimal model architecture, which is subsequently utilized for the training phase. However, some existing federated MR image reconstruction methods seldom consider inter-client fairness issues, leading to performance degradation of the trained global model on out-of-distribution data, especially when faced with statistical heterogeneity among clients. Intuitively, if the global model does not discriminate in performance on known clients (i.e., the performance gap is small on each client), then it logically should exhibit respectable performance even when faced with out-of-distribution data. Based on this notion, we introduce a fairness adjustment approach that utilizes the difference between the empirical risks of the global and local models in each communication round. This approach dynamically adjusts the aggregation weights of the clients, aiming to progressively diminish the performance gap of the global model across the known clients. As a result, this approach gradually improves the generalization performance of the global model on out-of-distribution data. Formally, the empirical risk gap between global and local models can be expressed as:

\begin{equation}
   \mathcal{G}_{D_{c}}(\theta^{r}) =\mathbf{\ell}_{D_{c}}(\theta^{r})-\mathbf{\ell}_{D_{c}}(\hat{\theta}_{c}^{r-1}), c=1, 2, \cdots, C
   \label{equ9}
\end{equation}
where $\ell_{D_{c}}(\theta^{r})$ and $\ell_{D_{c}}(\hat{\theta}_{c}^{r-1})$ denote the empirical risk of the global and local model, respectively. After completing local model training, the client sends the empirical risk gap $\mathcal{G}_{D_{c}}(\theta^{r}) $ and local model parameters $\hat{\theta}_{c}^{r}$ to the server. After the server receives $\mathcal{G}_{D_{c}}(\theta^{r}) $ and $\hat{\theta}_{c}^{r}$ from each client, it first updates the aggregation weight of each client according to the fairness adjustment approach, which is formulated as:

\begin{equation}
\begin{array}{l}
\begin{cases}
  \beta _{c}^{r}=a_{c}^{r-1}& \text{ if } \mathcal{G}_{D_{c}}(\theta^{r}) \le 0 \\
  \beta _{c}^{r}=a_{c}^{r-1}+\gamma\frac{\mathcal{G}_{D_{c}}(\theta^{r})}{\max(\mathcal{G}_{D_{\acute{c}}}(\theta^{r}))} & \text{ if } \mathcal{G}_{D_{c}}(\theta^{r})>0
\end{cases}
\end{array}
\label{equ10}
\end{equation}

\begin{equation}
  a_{c}^{r}=\frac{\beta_{c}^{r}}{ {\textstyle \sum_{c=1}^{C}} \beta _{c}^{r}}, \qquad \sum_{c=1}^{C}a_{c}^{r}=1
  \label{equ11}
\end{equation}
where $\acute{c}=\{1, 2, \cdots, C\}$, $a_{c}^{r-1}$ denotes the aggregation weight of the $c^{th}$ client at the $(r-1)^{th}$ communication round, $a_{c}^{r}$ is the weight updated by the fairness adjustment approach, $\gamma$ is utilized to control the magnitude of the weight increment. 

In each communication round, for each client $c$, if the empirical risk gap $\mathcal{G}_{D_{c}}(\theta^{r})\le 0$, we do not change its aggregation weight; if the empirical risk gap $\mathcal{G}_{D_{c}}(\theta^{r})>0$, which indicates that the global model does not generalize very well for local client, we increase its aggregation weight to reduce the empirical risk gap. In this way, we can continuously adjust the empirical risk gap on each client, and finally make the empirical risk gap smaller on each client, so that the global model can adapt to out-of-distribution data. The details of the training phase are shown in Algorithm. \ref{alg2}.

\section{Experiments}
\subsection{Experimental Setup}
\IEEEpubidadjcol  
\subsubsection{Datasets}
\begin{figure}[ht]
    \centering
    \setlength{\abovecaptionskip}{0.cm}
    \includegraphics[width=6cm, height=6cm, keepaspectratio]{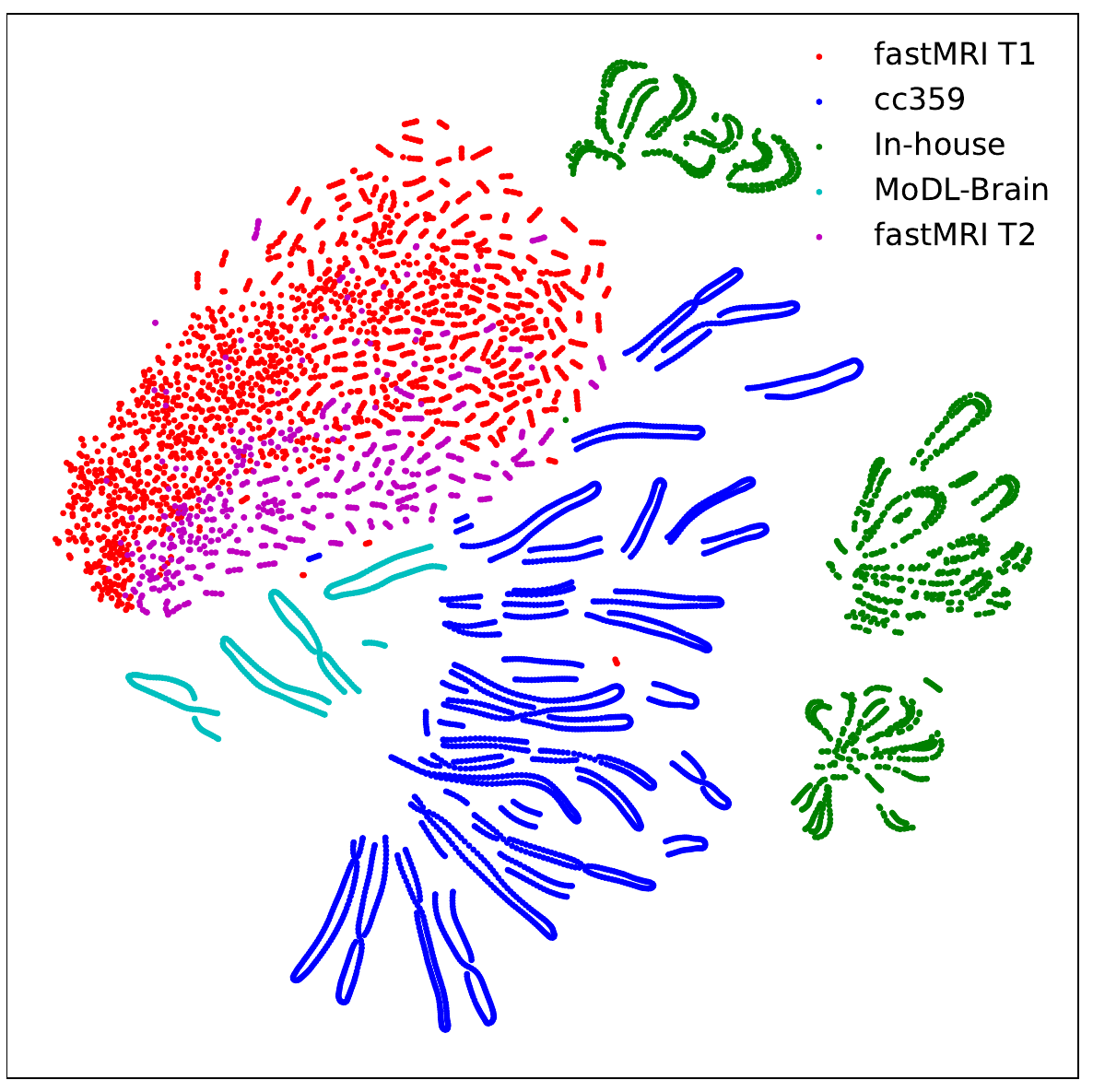}
    \caption{t-SNE visualizations of data distribution from four datasets: \textbf{fastMRI}, \textbf{cc359}, \textbf{MoDL-Brain}, and \textbf{In-house}.
    }
    \label{fig3}
\end{figure}

In this paper, we use three public datasets (\textbf{fastMRI}\cite{knoll2020fastmri}, \textbf{cc359}\footnote{https://sites.google.com/view/calgary-campinas-dataset/mr-reconstruction-\newline challenge}, \textbf{MoDL-Brain}\footnote{https://github.com/hkaggarwal/modl}) and one \textbf{In-house} dataset\cite{wang2020deepcomplexmri}. Details of the datasets are provided as follows: 1) \textbf{fastMRI}: a large public MRI dataset where we use only the T1 and T2 data of the brain, each including 285 subjects respectively. Among them, the T2 data is only used for validation experiments of different contrast; 2) \textbf{cc359}: there are 35 T1-weighted MR scans acquired on a clinical MR scanner (Discovery MR 750; General Electric (GE) Healthcare, Waukesha, WI); 3) \textbf{MoDL-Brain}: there are 524 slices in total, and this dataset is only used for validation experiments of unseen center; 4) \textbf{In-house}: There are T1, T2, and PD data for 22 subjects acquired in a United Imaging system, uMR 790. All data were divided into training, validation, and testing sets with the ratio of 7: 1: 2. Fig. \ref{fig3} shows the t-SNE visualization results of the distribution of the four datasets, from which it can be seen that the distribution of the four datasets is still quite different, which also shows the problem of statistical heterogeneity of the client. And the difference in data distribution between fastMRI T1 and T2 is somewhat smaller.

\subsubsection{Competing Methods}
To validate our proposed method, we compare it with six state-of-the-art FL algorithms, including: 1) FedAvg\cite{mcmahan2017fedavg}, a classical FL algorithm that is trained by average aggregation method; 2) FedProx\cite{li2020fedprox}, learns a global model by adding a regularization term to the local objective function; 3) MOON\cite{li2021moon}, an FL algorithm that corrects for local update through contrastive learning of model representations; 4) AM\cite{liu2021feddg}, a Fourier-based augmentation method, which was used in \cite{liu2021feddg}, we removed the segmentation task; 5) FedMRI\cite{feng2022fedmri}, a personalized FL algorithm that adds a weighted contrastive regularization term to each client;
6) FedALA\cite{zhang2023fedala}, learns personalized client model by local adaptive aggregation method.

\subsubsection{Implementation Details}
All the networks were trained using the PyTorch framework with one NVIDIA RTX A6000 GPU (with 48GB memory). The whole process is divided into two phases. In the searching phase, the Adam optimizer with a learning rate of 1e-4 is used to update architecture parameters, and the weight decay is set to 1e-3. Besides, the Adam optimizer with a learning rate of 1e-3 is used to update local model parameters. We set the batch size to 4. The models are trained with 50 global communication rounds and 5 local epochs for each round. In the training phase, the AdamW optimizer is adopted. We set the batch size and initial learning rate to 24 and 1e-3, respectively. Networks are trained with 150 global communication rounds and 5 local epochs for each round.

\subsection{In-Distribution Results}
To fully compare the performance of our method, we set up two scenarios: \textbf{In-Distribution} and \textbf{Out-of-Distribution} (different sampling pattern, acceleration rate, contrast, and unseen center) experiments. In this section, only the results of the \textbf{In-Distribution} scenario are presented, and the results of the \textbf{Out-of-Distribution} scenario will be presented in IV. C. The \textbf{In-Distribution} experiments are conducted using a 1D random 4$\times$ mask.

Table \ref{tab1} provides the quantitative results of the different methods under \textbf{In-Distribution} scenario. From Table \ref{tab1}, we can see that our method is superior to other FL methods in terms of PSNR and SSIM metrics. Compared with the classical FedAvg method, the average value of the PSNR metric of our GAutoMRI improves from 32.3288 dB to 35.5253 dB, and the average value of the SSIM metric increases from 0.8958 to 0.9384. This is primarily attributed to the fact that FedAvg uses an average aggregation approach, which does not adapt well to the statistical heterogeneity of the client. Comparatively, personalized federated learning methods (FedMRI and FedALA) perform better than FedAvg. They exhibit improved adaptability to local data distributions by maintaining client specificity. The \textbf{In-house} dataset comprises data with three different contrasts, resulting in greater heterogeneity, as illustrated in Fig. \ref{fig3}. Therefore, the performance of all methods on this dataset is slightly lower compared to their performance on the other two datasets. However, relatively speaking, our method still maintains a better performance, proving the superiority of our methods. Furthermore, our fairness adjustment approach contributes to achieving an obviously improved performance balance for each client.

\begin{figure*}[htbp]
    \centering
    \setlength{\abovecaptionskip}{0.cm}
    \includegraphics[width=14cm, height=10cm, keepaspectratio]{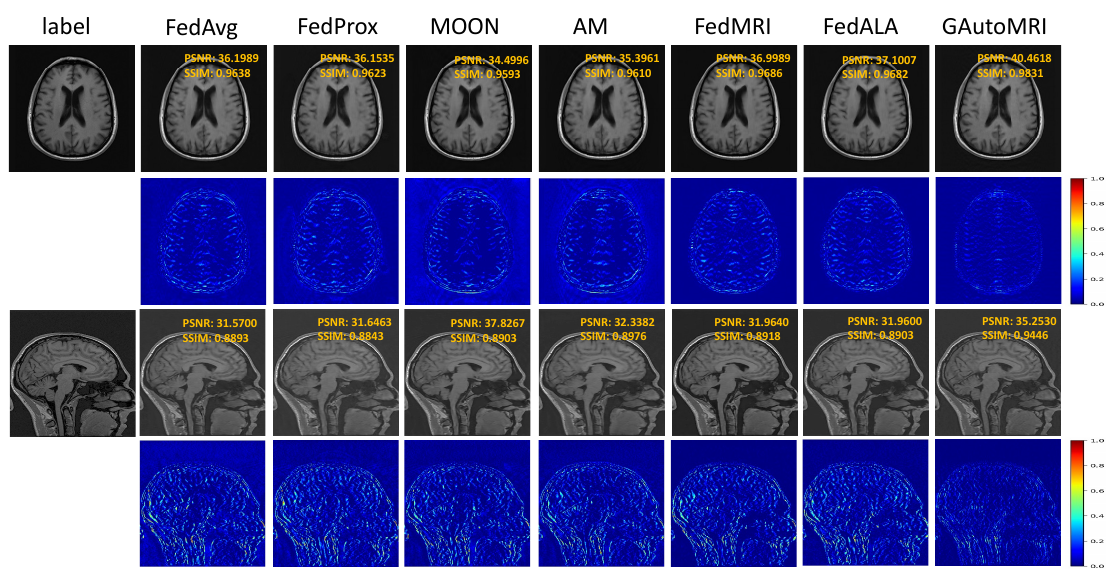}
    \caption{Qualitative reconstruction results of different methods under \textbf{In-Distribution} scenario. From left to right, the eight images corresponding to the reference image, the reconstructed images of FedAvg, FedProx, MOON, AM, FedMRI, FedALA, and our GAutoMRI, respectively. The second and fourth rows plot the corresponding error maps.
    }
    \label{fig4}
\end{figure*}

In addition to the quantitative results, the qualitative results of the different methods are also presented in Fig. \ref{fig4}. The reference images and reconstruction results from the comparative methods are displayed from left to right, with the corresponding error maps shown in the second and fourth rows. The corresponding quantitative metrics are also provided in the upper-right corner of the reconstructed images. From Fig. \ref{fig4}, our method obtains the best reconstruction quality and the smallest error, which also indicates that our method is superior to other methods.

\begin{table*}[ht]
\centering
\caption{Quantitative results of different methods with respect to \textbf{In-Distribution} scenario. Bold numbers indicate our results. Detailed analyses are described in III.B.}
\label{tab1}
\renewcommand{\arraystretch}{1.2} 
\resizebox{0.8\textwidth}{!}{
\tiny 
\begin{tabular}{cllllllll}
\toprule[1pt]
\multirow{3}{*}{method} & \multicolumn{8}{c}{In-Distribution}                                                                                                                                                                            \\ \cline{2-9} 
                        & \multicolumn{2}{c}{fastMRI T1}                      & \multicolumn{2}{c}{cc359}                           & \multicolumn{2}{c}{In-house}                        & \multicolumn{2}{c}{avg}                             \\ \cline{2-9} 
                        & \multicolumn{1}{c}{PSNR} & \multicolumn{1}{c}{SSIM} & \multicolumn{1}{c}{PSNR} & \multicolumn{1}{c}{SSIM} & \multicolumn{1}{c}{PSNR} & \multicolumn{1}{c}{SSIM} & \multicolumn{1}{c}{PSNR} & \multicolumn{1}{c}{SSIM} \\ 
\midrule[1pt]
FedAvg\cite{mcmahan2017fedavg}                  & 35.7327                  & 0.9551                   & 31.6436                  & 0.9060                   & 29.6101                  & 0.8262                   & 32.3288                  & 0.8958                   \\
FedProx\cite{li2020fedprox}                 & 35.7656                  & 0.9544                   & 31.6642                  & 0.9039                   & 29.9135                  & 0.8317                   & 32.4478                  & 0.8967                   \\
MOON\cite{li2021moon}                    & 35.4647                  & 0.9547                   & 31.7809                  & 0.9088                   & 26.0787                  & 0.7490                   & 31.1081                  & 0.8708                   \\
AM\cite{liu2021feddg}                      & 35.8460                  & 0.9559                   & 32.2660                  & 0.9125                   & 27.7329                  & 0.7926                   & 31.9483                  & 0.8870                   \\
FedMRI\cite{feng2022fedmri}                  & 36.3148                  & 0.9588                   & 31.7741                  & 0.9070                   & 30.7310                  & 0.8638                   & 32.9400                  & 0.9099                   \\
FedALA\cite{zhang2023fedala}                  & 36.5881                  & 0.9603                   & 31.9418                  & 0.9076                   & 31.0660                  & 0.8695                   & 33.1986                  & 0.9125                   \\ 
GAutoMRI                & \textbf{39.5964}                  & \textbf{0.9767}                   & \textbf{35.1023}                  & \textbf{0.9506}                   & \textbf{31.8779}                  & \textbf{0.8880}                   & \textbf{35.5253}                  & \textbf{0.9384}                   \\ 
\bottomrule[1pt]
\end{tabular}
}
\end{table*}

\subsection{Out-of-Distribution Results}
To demonstrate the performance of our proposed GAutoMRI on out-of-distribution data, we compare the model performance at different sampling pattern (1D random 4$\times$ for training; 1D equispaced 4$\times$ for testing), acceleration rate (1D random 4$\times$ for training; 1D random 6$\times$ for testing), contrast (fastMRI T2) and unseen center (\textbf{MoDL-Brain}). We divided it into two parts: 1) different sampling pattern and acceleration rate, and 2) different contrast and unseen center.

\subsubsection{Different Sampling Pattern and Acceleration Rate}
Table \ref{tab2} shows the quantitative results at different sampling pattern and acceleration rate. In terms of PSNR and SSIM values, our proposed method exhibits superior performance compared to the other methods. Compared to FedAvg, PSNR and SSIM improved from 27.8763 dB and 0.8271 to 30.0927 dB and 0.8602 under different sampling pattern; and PSNR and SSIM improved from 28.6256 dB and 0.8403 to 31.8182 dB and 0.8938 under different acceleration rate. This indicates that our approach demonstrates obviously improved performance on out-of-distribution data, and also shows that our approach has better generalization performance. The \textbf{In-house} dataset consists of three modalities, resulting in a notably heightened level of data heterogeneity. Consequently, the performance of all methods is expected to exhibit a slight reduction. However, the performance of our method remains the best. 

\begin{table}[ht]
\centering
\caption{Quantitative results of different methods with respect to \textbf{Out-of-Distribution} scenario (sampling pattern and acceleration rate). Bold numbers indicate our results. Detailed analyses are described in III.C.}
\label{tab2}
\renewcommand{\arraystretch}{1.2} 
\tiny 
\resizebox{0.5\textwidth}{!}{
\begin{tabular}{cllll}
\toprule[1pt]
\multirow{3}{*}{method} & \multicolumn{4}{c}{Out-of-Distribution}                                                                   \\ \cline{2-5} 
                        & \multicolumn{2}{c}{Sampling pattern}                & \multicolumn{2}{c}{Acceleration rate}               \\ \cline{2-5} 
                        & \multicolumn{1}{c}{PSNR} & \multicolumn{1}{c}{SSIM} & \multicolumn{1}{c}{PSNR} & \multicolumn{1}{c}{SSIM} \\ 
\midrule[1pt]
FedAvg\cite{mcmahan2017fedavg}                  & 27.8763                  & 0.8271                   & 28.6256                  & 0.8403                   \\
FedProx\cite{li2020fedprox}                 & 27.9759                  & 0.8279                   & 28.7533                  & 0.8425                   \\
MOON\cite{li2021moon}                    & 27.0243                  & 0.7992                   & 27.7452                  & 0.8134                   \\
AM\cite{liu2021feddg}                      & 27.7492                  & 0.8226                   & 28.7145                  & 0.8401                   \\
FedMRI\cite{feng2022fedmri}                  & 28.1829                  & 0.8438                   & 28.8346                  & 0.8531                   \\
FedALA\cite{zhang2023fedala}                  & 28.0714                  & 0.8470                   & 28.9034                  & 0.8526                   \\ 
GAutoMRI                & \textbf{30.0927}                  & \textbf{0.8602}                   & \textbf{31.8182}                  & \textbf{0.8938}                   \\ 
\bottomrule[1pt]
\end{tabular}
}
\end{table}

\begin{figure*}[t]
    \centering
    \setlength{\abovecaptionskip}{0.cm}
    \includegraphics[width=14cm, height=8cm, keepaspectratio]{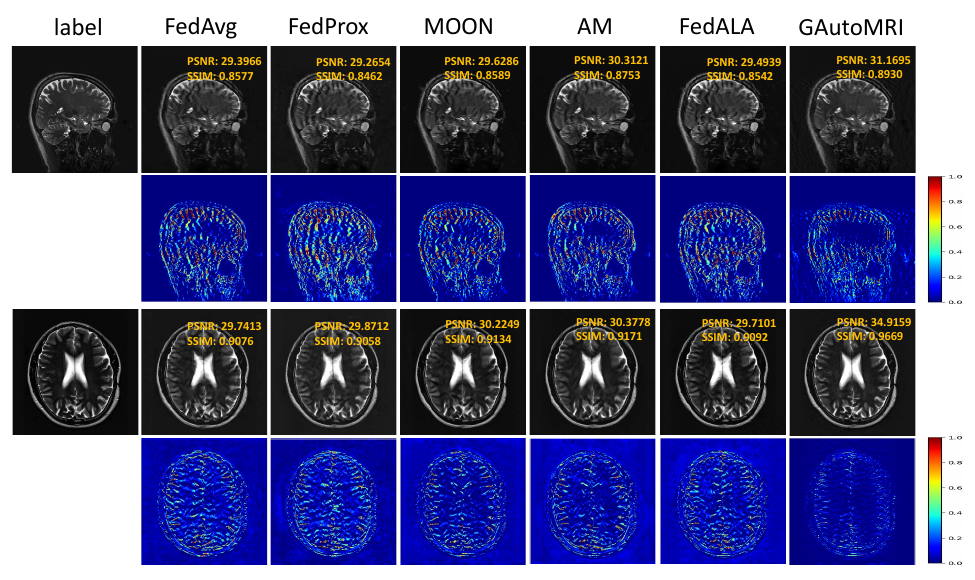}
    \caption{Qualitative reconstruction results of different methods under \textbf{Out-of-Distribution} scenario (different contrast and unseen center). From left to right, the seven images corresponding to the reference image, the reconstructed images of FedAvg, FedProx, MOON, AM, FedALA, and our GAutoMRI, respectively. The second and fourth rows plot the corresponding error maps.
    }
    \label{fig5}
\end{figure*}

\subsubsection{Different Contrast and Unseen Center}
Table \ref{tab3} shows the quantitative results of the generalization performance at different contrast and unseen center, and the experimental results show that the performance of our method is the best. Compared to FedAvg, under different contrast, PSNR improved from 29.7607 dB to 34.6154 dB, and SSIM improved from 0.9126 to 0.9647. Under the unseen center, PSNR improved from 29.0923 dB to 31.5570 dB, and SSIM improved from 0.7879 to 0.8834. As depicted in Fig. \ref{fig3}, the disparity between the data distributions of fastMRI T1 and T2 is relatively small, resulting in somewhat lesser performance degradation for all methods in scenarios involving different contrast. Generalizing models trained on heterogeneous data distributions to the unseen center is a more challenging task. However, our GAutoMRI also maintains a better performance, attributed to the fairness adjustment approach we introduced. This approach enhances generalization under the unseen center by mitigating the performance gap of the global model across known centers. AM is a domain generalization method, and its performance is the best except for our method. Furthermore, we also compared the parameter size of individual client models. As indicated in Table \ref{tab3}, the model obtained through the neural architecture search exhibits a notably reduced parameter size, which indicates that our approach is more efficient. In addition, we visualized the reconstructed images and the corresponding error maps for the different methods, as shown in Fig. \ref{fig5}. From the error maps, it can be seen that the error of our method is minimized, indicating our method outperforms the other methods. Overall, our proposed method has better generalization performance.

\begin{table}[t]
\centering
\caption{Quantitative results of different methods with respect to \textbf{Out-of-Distribution} scenario (different contrast and unseen center). Bold numbers indicate our results. Detailed analyses are described in III.C.}
\label{tab3}
\renewcommand{\arraystretch}{1.2} 
\scriptsize 
\resizebox{0.5\textwidth}{!}{
\begin{tabular}{cccccc}
\toprule[1.5pt]
\multirow{3}{*}{method} & \multicolumn{5}{c}{Out-of-Distribution}                                                                  \\ \cline{2-6} 
                        & \multicolumn{2}{c}{Different contrast} & \multicolumn{2}{c}{Unseen Center} & \multirow{2}{*}{\#Params/M} \\ \cline{2-5}
                        & PSNR               & SSIM              & PSNR            & SSIM            &                             \\
\midrule[1.5pt]
FedAvg\cite{mcmahan2017fedavg}                  & 29.7607            & 0.9126            & 29.0923         & 0.7879          & 7.76                        \\
FedProx\cite{li2020fedprox}                 & 29.8297            & 0.9107            & 29.0932         & 0.7840          & 7.76                        \\
MOON\cite{li2021moon}                    & 30.0428            & 0.9174            & 29.2461         & 0.7891          & 41.44                       \\
AM\cite{liu2021feddg}                      & 30.3382            & 0.9206            & 30.1058         & 0.8168          & 7.76                        \\
FedALA\cite{zhang2023fedala}                  & 29.7476            & 0.9132            & 29.0459         & 0.7842          & 7.76                        \\ 
GAutoMRI                & \textbf{34.6154}            & \textbf{0.9647}            & \textbf{31.5570}         & \textbf{0.8834}         & \textbf{0.0487}                      \\ 
\bottomrule[1.5pt]
\end{tabular}
}
\end{table}

\subsection{Ablation Studies}
In this section, we conduct ablation studies and utilize a radar chart to analyze the performance differences among different methods within the source domain.

\subsubsection{Ablation Studies of Fairness Adjustment Approach}
Table \ref{tab4} shows the quantitative results of the ablation studies conducted on the fairness adjustment (FA) approach under the \textbf{In-Distribution} scenario. where \textit{w/o}(FA) denotes the result without the fairness adjustment approach. Compared to the hand-designed model, the performance of the model obtained using the neural architecture search exhibits an improved performance. The PSNR and SSIM metrics have been improved from 33.8465 dB and 0.9126 to 35.2859 dB and 0.9342, respectively. GAutoMRI requires only about 18\% of the computational and parameter sizes of the baseline model, underscoring the efficiency of the searched model. Upon incorporating the fairness adjustment approach, there is a clear improvement in the average metrics, even though there is a slight decline in performance on the \textbf{cc359} dataset. This is because increasing the aggregation weight of one client may affect the aggregation weight of other clients, resulting in slight performance degradation, which is acceptable. 

\begin{figure}[htbp]
    \centering
    \setlength{\abovecaptionskip}{0.cm}
    \includegraphics[width=8cm, height=8cm, keepaspectratio]{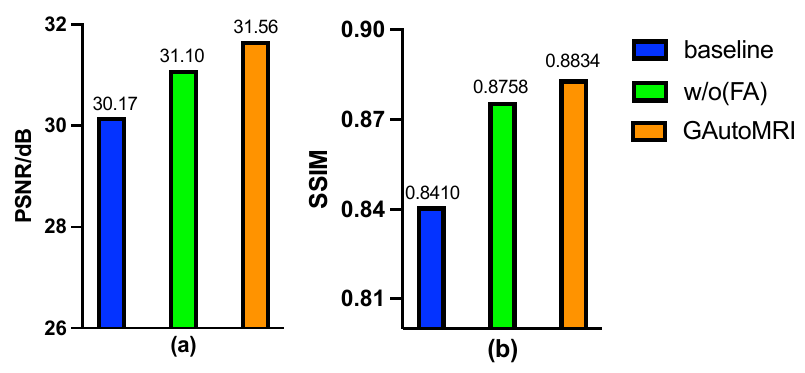}
    \caption{Quantitative results of ablation experiment under \textbf{Out-of-Distribution} scenario (unseen center).
    }
    \label{fig6}
\end{figure}

\begin{figure}[htbp]
    \centering
    \setlength{\abovecaptionskip}{0.cm}
    \includegraphics[width=8cm, height=8cm, keepaspectratio]{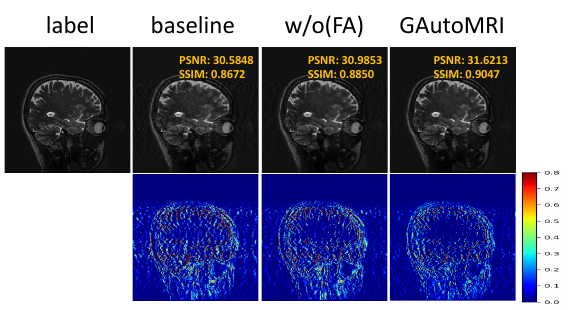}
    \caption{Qualitative results of ablation experiment under \textbf{Out-of-Distribution} scenario (unseen center).
    }
    \label{fig7}
\end{figure}

In addition, we explored the role of FA on the generalization performance of the model. Figs. \ref{fig6} and \ref{fig7} show the quantitative and qualitative results of FA on unseen center, respectively. From the quantitative metrics, when the fairness adjustment approach is added, the performance on the unseen center is improved, which verifies the facilitating effect of FA on improving the generalization of the model. From the error maps of the qualitative results, when FA is introduced, the error of the reconstructed image is reduced. Overall, the fairness adjustment approach promotes the generalization performance of the model.

\begin{table*}[ht]
\centering
\caption{Ablation study about fairness adjustment approach under \textbf{In-Distribution} scenario (\textit{without (w/o)}).}
\label{tab4}
\renewcommand{\arraystretch}{1.2} 
\resizebox{0.8\textwidth}{!}{
\normalsize 
\begin{tabular}{ccccccccccc}
\toprule[1pt]
\multirow{2}{*}{method} & \multicolumn{2}{c}{fastMRI T1} & \multicolumn{2}{c}{cc359} & \multicolumn{2}{c}{In-house} & \multicolumn{2}{c}{avg} & \multirow{2}{*}{FLOPs/G} & \multirow{2}{*}{\#Param/M} \\ \cline{2-9}
                        & PSNR           & SSIM          & PSNR         & SSIM       & PSNR          & SSIM         & PSNR        & SSIM      &                          &                             \\
\midrule[1pt]
baseline                & 37.4104        & 0.9598        & 33.8985      & 0.9320     & 30.2305       & 0.8461       & 33.8465     & 0.9126    & 85.6503                  & 0.2614                      \\ 
\textit{w/o}(FA)                 & 39.5233        & 0.9778        & 35.1219      & 0.9520     & 31.2125       & 0.8729       & 35.2859     & 0.9342    & 15.7444                  & 0.0487                      \\ 
GAutoMRI                & 39.5964        & 0.9767        & 35.1023      & 0.9506     & 31.8779       & 0.8880       & 35.5255     & 0.9384    & -                        & -                           \\
\bottomrule[1pt]
\end{tabular}
}
\end{table*}

\subsubsection{Testing Errors in Source Domain}
\begin{figure}[htbp]
    \centering
    \setlength{\abovecaptionskip}{0.cm}
    \includegraphics[width=7cm, height=7cm, keepaspectratio]{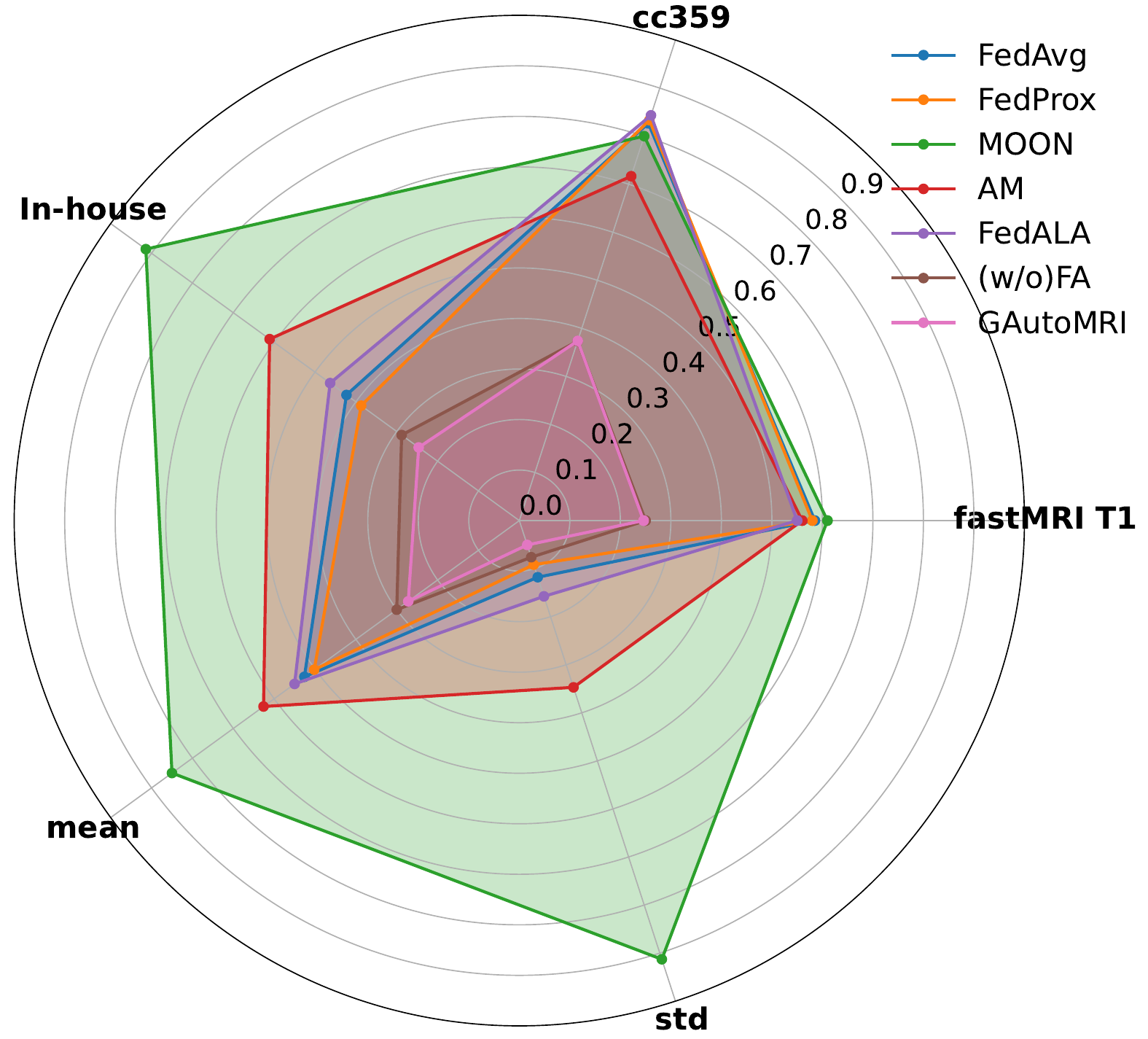}
    \caption{Radar chart of the testing errors of the global model under \textbf{In-Distribution} scenario.
    }
    \label{fig8}
\end{figure}

This section focuses on analyzing the fairness of the different methods, and Fig. \ref{fig8} shows a radar chart of the testing errors of the different methods under \textbf{In-Distribution} scenario. The chart features five axes that symbolize the testing errors of each method across the three datasets, as well as the mean and standard deviation of these testing errors. The magnitude of the standard deviation is utilized to portray the fairness of the model, as depicted in Fig. \ref{fig8}, the standard deviation of our method is the smallest and the area enclosed by the pentagon is also the smallest, indicating the better fairness of our proposed GAutoMRI. The contribution of FA to model fairness can also be observed.

\section{Conclusion}
In this paper, we focus on addressing the generalization issue of deep learning MR image reconstruction models towards heterogeneous data captured by different scanning devices and imaging protocols,  using multi-institutional data in a privacy-preserving manner. Specifically, we propose a generalizable federated neural architecture search framework, GAutoMRI. In the framework, we first design an adapted search space for undersampled MR image reconstruction task, and then learn an optimal model from heterogeneous data distributions with differentiable federated neural architecture search. In addition, we design a fairness adjustment approach that
can enable the model to learn features fairly from inconsistent
distributions of different devices.  Extensive experiments on different sampling patterns, acceleration rates, contrasts, and unseen centers demonstrate that our proposed GAutoMRI has better generalization and performances compared to six state-of-the-art federated learning methods.

\section*{Acknowledgments}
This research was partly supported by the National Natural Science Foundation of China (62222118, U22A2040), Guangdong Provincial Key Laboratory of Artificial Intelligence in Medical Image Analysis and Application (2022B1212010011), Shenzhen Science and Technology Program (RCYX20210706092104034, JCYJ20220531100213029), Key Laboratory for Magnetic Resonance and Multimodality Imaging of Guangdong Province (2023B1212060052), and the Major Key Project of Peng Cheng Laboratory (PCL2023AS1-2).

\bibliographystyle{IEEEtran}
\bibliography{ref}

\begin{thebibliography}{10}
\providecommand{\url}[1]{#1}
\csname url@samestyle\endcsname
\providecommand{\newblock}{\relax}
\providecommand{\bibinfo}[2]{#2}
\providecommand{\BIBentrySTDinterwordspacing}{\spaceskip=0pt\relax}
\providecommand{\BIBentryALTinterwordstretchfactor}{4}
\providecommand{\BIBentryALTinterwordspacing}{\spaceskip=\fontdimen2\font plus
\BIBentryALTinterwordstretchfactor\fontdimen3\font minus
  \fontdimen4\font\relax}
\providecommand{\BIBforeignlanguage}[2]{{%
\expandafter\ifx\csname l@#1\endcsname\relax
\typeout{** WARNING: IEEEtran.bst: No hyphenation pattern has been}%
\typeout{** loaded for the language `#1'. Using the pattern for}%
\typeout{** the default language instead.}%
\else
\language=\csname l@#1\endcsname
\fi
#2}}
\providecommand{\BIBdecl}{\relax}
\BIBdecl

\bibitem{korkmaz2022unsupervised}
Y.~Korkmaz, S.~U. Dar, M.~Yurt, M.~{\"O}zbey, and T.~Cukur, ``Unsupervised
  {MRI} reconstruction via zero-shot learned adversarial transformers,''
  \emph{IEEE Transactions on Medical Imaging}, vol.~41, no.~7, pp. 1747--1763,
  2022.

\bibitem{sandino2020compressed}
C.~M. Sandino, J.~Y. Cheng, F.~Chen, M.~Mardani, J.~M. Pauly, and S.~S.
  Vasanawala, ``Compressed sensing: From research to clinical practice with
  deep neural networks: Shortening scan times for magnetic resonance imaging,''
  \emph{IEEE signal processing magazine}, vol.~37, no.~1, pp. 117--127, 2020.

\bibitem{wang2016accelerating}
S.~Wang, Z.~Su, L.~Ying, X.~Peng, S.~Zhu, F.~Liang, D.~Feng, and D.~Liang,
  ``Accelerating magnetic resonance imaging via deep learning,'' in \emph{2016
  IEEE 13th international symposium on biomedical imaging (ISBI)}.\hskip 1em
  plus 0.5em minus 0.4em\relax IEEE, 2016, pp. 514--517.

\bibitem{hammernik2018learning}
K.~Hammernik, T.~Klatzer, E.~Kobler, M.~P. Recht, D.~K. Sodickson, T.~Pock, and
  F.~Knoll, ``Learning a variational network for reconstruction of accelerated
  {MRI} data,'' \emph{Magnetic resonance in medicine}, vol.~79, no.~6, pp.
  3055--3071, 2018.

\bibitem{mardani2018deep}
M.~Mardani, E.~Gong, J.~Y. Cheng, S.~S. Vasanawala, G.~Zaharchuk, L.~Xing, and
  J.~M. Pauly, ``Deep generative adversarial neural networks for compressive
  sensing {MRI},'' \emph{IEEE transactions on medical imaging}, vol.~38, no.~1,
  pp. 167--179, 2018.

\bibitem{yang2018admm}
Y.~Yang, J.~Sun, H.~Li, and Z.~Xu, ``{ADMM-CSNet}: A deep learning approach for
  image compressive sensing,'' \emph{IEEE transactions on pattern analysis and
  machine intelligence}, vol.~42, no.~3, pp. 521--538, 2018.

\bibitem{quan2018compressed}
T.~M. Quan, T.~Nguyen-Duc, and W.-K. Jeong, ``Compressed sensing {MRI}
  reconstruction using a generative adversarial network with a cyclic loss,''
  \emph{IEEE transactions on medical imaging}, vol.~37, no.~6, pp. 1488--1497,
  2018.

\bibitem{aggarwal2018modl}
H.~K. Aggarwal, M.~P. Mani, and M.~Jacob, ``{MoDL}: Model-based deep learning
  architecture for inverse problems,'' \emph{IEEE transactions on medical
  imaging}, vol.~38, no.~2, pp. 394--405, 2018.

\bibitem{zhu2018image}
B.~Zhu, J.~Z. Liu, S.~F. Cauley, B.~R. Rosen, and M.~S. Rosen, ``Image
  reconstruction by domain-transform manifold learning,'' \emph{Nature}, vol.
  555, no. 7697, pp. 487--492, 2018.

\bibitem{qin2018convolutional}
C.~Qin, J.~Schlemper, J.~Caballero, A.~N. Price, J.~V. Hajnal, and D.~Rueckert,
  ``Convolutional recurrent neural networks for dynamic {MR} image
  reconstruction,'' \emph{IEEE transactions on medical imaging}, vol.~38,
  no.~1, pp. 280--290, 2018.

\bibitem{han2018deep}
Y.~Han, J.~Yoo, H.~H. Kim, H.~J. Shin, K.~Sung, and J.~C. Ye, ``Deep learning
  with domain adaptation for accelerated projection-reconstruction {MR},''
  \emph{Magnetic resonance in medicine}, vol.~80, no.~3, pp. 1189--1205, 2018.

\bibitem{akccakaya2019scan}
M.~Ak{\c{c}}akaya, S.~Moeller, S.~Weing{\"a}rtner, and K.~U{\u{g}}urbil,
  ``Scan-specific robust artificial-neural-networks for k-space interpolation
  ({RAKI}) reconstruction: Database-free deep learning for fast imaging,''
  \emph{Magnetic resonance in medicine}, vol.~81, no.~1, pp. 439--453, 2019.

\bibitem{wang2021deep}
S.~Wang, T.~Xiao, Q.~Liu, and H.~Zheng, ``Deep learning for fast {MR} imaging:
  a review for learning reconstruction from incomplete k-space data,''
  \emph{Biomedical Signal Processing and Control}, vol.~68, p. 102579, 2021.

\bibitem{lee2018deep}
D.~Lee, J.~Yoo, S.~Tak, and J.~C. Ye, ``Deep residual learning for accelerated
  {MRI} using magnitude and phase networks,'' \emph{IEEE Transactions on
  Biomedical Engineering}, vol.~65, no.~9, pp. 1985--1995, 2018.

\bibitem{wang2020deepcomplexmri}
S.~Wang, H.~Cheng, L.~Ying, T.~Xiao, Z.~Ke, H.~Zheng, and D.~Liang,
  ``{DeepcomplexMRI}: Exploiting deep residual network for fast parallel {MR}
  imaging with complex convolution,'' \emph{Magnetic Resonance Imaging},
  vol.~68, pp. 136--147, 2020.

\bibitem{eo2018kiki}
T.~Eo, Y.~Jun, T.~Kim, J.~Jang, H.-J. Lee, and D.~Hwang, ``{KIKI}-net:
  cross-domain convolutional neural networks for reconstructing undersampled
  magnetic resonance images,'' \emph{Magnetic resonance in medicine}, vol.~80,
  no.~5, pp. 2188--2201, 2018.

\bibitem{liang2020deep}
D.~Liang, J.~Cheng, Z.~Ke, and L.~Ying, ``Deep magnetic resonance image
  reconstruction: Inverse problems meet neural networks,'' \emph{IEEE Signal
  Processing Magazine}, vol.~37, no.~1, pp. 141--151, 2020.

\bibitem{kaissis2020secure}
G.~A. Kaissis, M.~R. Makowski, D.~R{\"u}ckert, and R.~F. Braren, ``Secure,
  privacy-preserving and federated machine learning in medical imaging,''
  \emph{Nature Machine Intelligence}, vol.~2, no.~6, pp. 305--311, 2020.

\bibitem{kaissis2021end}
G.~Kaissis, A.~Ziller, J.~Passerat-Palmbach, T.~Ryffel, D.~Usynin, A.~Trask,
  I.~Lima~Jr, J.~Mancuso, F.~Jungmann, M.-M. Steinborn \emph{et~al.},
  ``End-to-end privacy preserving deep learning on multi-institutional medical
  imaging,'' \emph{Nature Machine Intelligence}, vol.~3, no.~6, pp. 473--484,
  2021.

\bibitem{li2020federated}
T.~Li, A.~K. Sahu, A.~Talwalkar, and V.~Smith, ``Federated learning:
  Challenges, methods, and future directions,'' \emph{IEEE signal processing
  magazine}, vol.~37, no.~3, pp. 50--60, 2020.

\bibitem{mcmahan2017fedavg}
B.~McMahan, E.~Moore, D.~Ramage, S.~Hampson, and B.~A. y~Arcas,
  ``Communication-efficient learning of deep networks from decentralized
  data,'' in \emph{Artificial intelligence and statistics}.\hskip 1em plus
  0.5em minus 0.4em\relax PMLR, 2017, pp. 1273--1282.

\bibitem{li2020fedprox}
T.~Li, A.~K. Sahu, M.~Zaheer, M.~Sanjabi, A.~Talwalkar, and V.~Smith,
  ``Federated optimization in heterogeneous networks,'' \emph{Proceedings of
  Machine learning and systems}, vol.~2, pp. 429--450, 2020.

\bibitem{li2021moon}
Q.~Li, B.~He, and D.~Song, ``Model-contrastive federated learning,'' in
  \emph{Proceedings of the IEEE/CVF Conference on Computer Vision and Pattern
  Recognition}, 2021, pp. 10\,713--10\,722.

\bibitem{liu2021feddg}
Q.~Liu, C.~Chen, J.~Qin, Q.~Dou, and P.-A. Heng, ``Feddg: Federated domain
  generalization on medical image segmentation via episodic learning in
  continuous frequency space,'' in \emph{Proceedings of the IEEE/CVF Conference
  on Computer Vision and Pattern Recognition}, 2021, pp. 1013--1023.

\bibitem{zhang2023fedala}
J.~Zhang, Y.~Hua, H.~Wang, T.~Song, Z.~Xue, R.~Ma, and H.~Guan, ``Fedala:
  Adaptive local aggregation for personalized federated learning,'' in
  \emph{Proceedings of the AAAI Conference on Artificial Intelligence},
  vol.~37, no.~9, 2023, pp. 11\,237--11\,244.

\bibitem{guo2021fl-mrcm}
P.~Guo, P.~Wang, J.~Zhou, S.~Jiang, and V.~M. Patel, ``Multi-institutional
  collaborations for improving deep learning-based magnetic resonance image
  reconstruction using federated learning,'' in \emph{Proceedings of the
  IEEE/CVF Conference on Computer Vision and Pattern Recognition}, 2021, pp.
  2423--2432.

\bibitem{feng2022fedmri}
C.-M. Feng, Y.~Yan, S.~Wang, Y.~Xu, L.~Shao, and H.~Fu,
  ``Specificity-preserving federated learning for {MR} image reconstruction,''
  \emph{IEEE Transactions on Medical Imaging}, 2022.

\bibitem{elmas2022federated}
G.~Elmas, S.~U. Dar, Y.~Korkmaz, E.~Ceyani, B.~Susam, M.~Ozbey, S.~Avestimehr,
  and T.~{\c{C}}ukur, ``Federated learning of generative image priors for {MRI}
  reconstruction,'' \emph{IEEE Transactions on Medical Imaging}, 2022.

\bibitem{wang2021federated}
Z.~Wang, X.~Fan, J.~Qi, C.~Wen, C.~Wang, and R.~Yu, ``Federated learning with
  fair averaging,'' in \emph{Proceedings of International Joint Conference on
  Artificial Intelligence}, 2021.

\bibitem{elsken2019neural}
T.~Elsken, J.~H. Metzen, and F.~Hutter, ``Neural architecture search: A
  survey,'' \emph{The Journal of Machine Learning Research}, vol.~20, no.~1,
  pp. 1997--2017, 2019.

\bibitem{yan2020neural}
J.~Yan, S.~Chen, Y.~Zhang, and X.~Li, ``Neural architecture search for
  compressed sensing magnetic resonance image reconstruction,''
  \emph{Computerized Medical Imaging and Graphics}, vol.~85, p. 101784, 2020.

\bibitem{huang2020enhanced}
Q.~Huang, D.~Yang, Y.~Xian, P.~Wu, J.~Yi, H.~Qu, and D.~Metaxas, ``Enhanced
  {MRI} reconstruction network using neural architecture search,'' in
  \emph{Machine Learning in Medical Imaging: 11th International Workshop, MLMI
  2020, Held in Conjunction with MICCAI 2020, Lima, Peru, October 4, 2020,
  Proceedings 11}.\hskip 1em plus 0.5em minus 0.4em\relax Springer, 2020, pp.
  634--643.

\bibitem{knoll2020fastmri}
F.~Knoll, J.~Zbontar, A.~Sriram, M.~J. Muckley, M.~Bruno, A.~Defazio,
  M.~Parente, K.~J. Geras, J.~Katsnelson, H.~Chandarana \emph{et~al.},
  ``{fastMRI}: A publicly available raw k-space and dicom dataset of knee
  images for accelerated {MR} image reconstruction using machine learning,''
  \emph{Radiology: Artificial Intelligence}, vol.~2, no.~1, p. e190007, 2020.

\bibitem{bello2017neural}
I.~Bello, B.~Zoph, V.~Vasudevan, and Q.~V. Le, ``Neural optimizer search with
  reinforcement learning,'' in \emph{International Conference on Machine
  Learning}.\hskip 1em plus 0.5em minus 0.4em\relax PMLR, 2017, pp. 459--468.

\bibitem{pham2018efficient}
H.~Pham, M.~Guan, B.~Zoph, Q.~Le, and J.~Dean, ``Efficient neural architecture
  search via parameters sharing,'' in \emph{International conference on machine
  learning}.\hskip 1em plus 0.5em minus 0.4em\relax PMLR, 2018, pp. 4095--4104.

\bibitem{yang2020cars}
Z.~Yang, Y.~Wang, X.~Chen, B.~Shi, C.~Xu, C.~Xu, Q.~Tian, and C.~Xu, ``Cars:
  Continuous evolution for efficient neural architecture search,'' in
  \emph{Proceedings of the IEEE/CVF Conference on Computer Vision and Pattern
  Recognition}, 2020, pp. 1829--1838.

\bibitem{sun2020automatically}
Y.~Sun, B.~Xue, M.~Zhang, G.~G. Yen, and J.~Lv, ``Automatically designing {CNN}
  architectures using the genetic algorithm for image classification,''
  \emph{IEEE transactions on cybernetics}, vol.~50, no.~9, pp. 3840--3854,
  2020.

\bibitem{liu2018darts}
H.~Liu, K.~Simonyan, and Y.~Yang, ``{DARTS}: Differentiable architecture
  search,'' in \emph{International Conference on Learning Representations},
  2018.

\bibitem{he2020milenas}
C.~He, H.~Ye, L.~Shen, and T.~Zhang, ``Milenas: Efficient neural architecture
  search via mixed-level reformulation,'' in \emph{Proceedings of the IEEE/CVF
  Conference on Computer Vision and Pattern Recognition}, 2020, pp.
  11\,993--12\,002.

\bibitem{he2020towards}
C.~He, M.~Annavaram, and S.~Avestimehr, ``Towards non-iid and invisible data
  with {FedNAS}: Federated deep learning via neural architecture search,'' in
  \emph{CVPR 2020 workshop on neural architecture search and beyond for
  representation learning}, 2020.

\end{thebibliography}

\newpage

 




\vfill

\end{document}